\documentclass[%
 reprint,
 amsmath,amssymb,
 aps,
]{revtex4-2}

\usepackage{graphicx}
\usepackage{dcolumn}
\usepackage{bm}

\begin{document}

\preprint{APS/123-QED}

\title{Clathrate Structure of Fullerite $\mathrm{C}_{60}$ }

\author{Jorge Laranjeira}
 \email{jorgelaranjeira@ua.pt}
 \altaffiliation{Departamento de Física and CICECO, Universidade de Aveiro, 3810-193 Aveiro, Portugal}
\author{Leonel Marques}%
\affiliation{Departamento de Física and CICECO, Universidade de Aveiro, 3810-193 Aveiro, Portugal}%

\author{Manuel Melle-Franco}
\affiliation{Departamento de Química and CICECO, Universidade de Aveiro, 3810-193 Aveiro, Portugal}%

\author{Karol Strutyński}
\affiliation{Departamento de Química and CICECO, Universidade de Aveiro, 3810-193 Aveiro, Portugal}%

\author{Manuel Barroso}%
\affiliation{Departamento de Física and I3N, Universidade de Aveiro, 3810-193 Aveiro, Portugal}%

\date{\today}

\begin{abstract}
Investigations of clathrate structures have gained a new impetus with the recent discovery of room-temperature superconductivity in metal hydrides. Here we report the finding, through density functional theory calculations, of a clathrate phase in the fullerite $\mathrm{C}_{60}$ system. Intermolecular bonds of the type 5/5 2+3 cycloaddition are induced between each $\mathrm{C}_{60}$ molecule and its twelve nearest neighbors in the face centered cubic lattice. Remarkably, this bonding creates on octahedral sites new $\mathrm{C}_{60}$ cages, identical to the original ones, and on tetrahedral sites distorted sodalite-like cages. The resulting carbon clathrate has a $\mathrm{Pm}\bar{3}$ simple cubic structure with half of the original face centered lattice constant. Eighty per cent of its atoms are $\mathrm{sp}^{3}$-hybridized, driving a narrow-gap semiconducting behavior, a moderate bulk modulus of 268GPa and an estimated hardness of 21.6GPa. This new phase is likely to be prepared by subjecting $\mathrm{C}_{60}$ to high pressure and high temperature conditions.
\end{abstract}

\maketitle



Clathrates are open-framework structures with face-sharing polyhedral cages that can contain guest species, atoms or molecules \cite{gashydrysdescsr,geclatrateexp,yamanakaclatrato,sun_hyd,dias_hyd,hemley_hyd,eremets_hyd,pickard_hydrides,strobel2020,Zhu2020,tao2015,liu2017cla,li2016_base,blasesupercond,blase2004,damien2010,sanmiguelc60clatratos}. Hydrates are an important group of clathrate structures, being the focus of intense investigations on the possibility of greenhouse gases storage \cite{gashydrysdescsr}. Group IV elemental solids can also form clathrates and, in particular, silicon and germanium clathrates have been synthesized in both guest and guest-free forms \cite{geclatrateexp,yamanakaclatrato}. Their properties can be tuned by inserting or exchanging guest atoms inside the polyhedral cages \cite{blasesupercond}. The stunning recent discovery of room-temperature superconductivity at high pressure in metal hydrides, in which the metal atoms are located inside the hydrogen cages, has fueled the investigations on clathrate structures \cite{sun_hyd,dias_hyd,hemley_hyd,eremets_hyd,pickard_hydrides}. 

Carbon has exceptional flexibility as shown by the profusion of allotropes with distinct physical and chemical properties \cite{sanmiguelc60clatratos,SUNDQVIST2021,prlwang2018,laran2018,li2016_base,blasesupercond,blase2004,damien2010,strout1993dim,dimondmech,yamanakaclatrato,endo,yamaka_sp3,lyapin2019,soldatov2020,mezouarprb2003,reviewpei}. Despite the large number of geometries carbon constructs may adopt, no carbon clathrate has been synthesized so far, in contrast to other group IV elements. There have been several theoretical studies discussing the stability of different carbon clathrates, such as $\mathrm{C}_{46}$, $\mathrm{C}_{34}$ and $\mathrm{C}_{24}$, whose construction is based on polyhedral cages and arrangements exhibited by clathrates of other chemical elements \cite{li2016_base,blasesupercond,blase2004,damien2010}. A significant step towards the synthesis of carbon clathrates has been done recently with the high-pressure synthesis of mixed carbon-boron clathrates, where metal atoms, strontium or lanthanum, are trapped inside ordered sodalite-like cages formed by carbon and boron \cite{strobel2020,Zhu2020}.

Notable candidates for the realization of carbon clathrates are some of the three-dimensional (3D) $\mathrm{C}_{60}$ polymers prepared at high pressure and high temperature. However, the crystal structures reported so far involve isolated $\mathrm{C}_{60}$ cages bonded to each other through $\mathrm{sp}^{3}$ carbons \cite{yamanakaclatrato}. Here we show, through density functional theory (DFT) calculations, that indeed fullerite $\mathrm{C}_{60}$ can also form an all-carbon clathrate structure through 3D polymerization of the molecules in the face centered cubic (fcc) lattice that creates two new polyhedral cages, sharing faces with each other and with the original $\mathrm{C}_{60}$ molecules. This carbon clathrate structure is statically and dynamically stable, being somewhat less stable than the known 3D $\mathrm{C}_{60}$ polymers, and is a narrow-gap semiconductor with a predicted indirect gap of 0.68eV. Despite a high ratio, eighty percent, of $\mathrm{sp}^{3}$ carbons, it displays moderate elastic moduli and a moderate hardness of 21.62GPa, probably a consequence of the elastic flexibility of the big $\mathrm{C}_{60}$ cages.


DFT calculations were performed using the Vienna ab initio simulation package (VASP) \cite{i32}. Perdew-Burke-Ernzerhof generalized gradient approximation was used to calculate the exchange-correlation energy \cite{i2}. Electron core interaction was described by the projector augmented wave pseudo-potentials supplied with VASP. The self-consistent-field process was stopped once the energy difference between two consecutive steps was smaller than $10^{-5}$eV; the k-points were converged within $10^{-5}$meV/atom. To achieve this convergence 6x6x6 k-point mesh was employed in the optimization of the atomic positions and lattice parameters. A gaussian smearing with standard deviation of 0.2 and an energy cutoff of 520eV were also used for the optimization.

Electron Density of States (DOS) calculations were performed using a 16×16×16 K-point grid. An energy cutoff of 400eV was employed and the tetrahedron method with Bloch corrections was used for electron smearing. For the electron band structure calculation, the same energy cutoff was used with the calculation being made using 20 K-points between the high symmetric K-points of the Brillouin zone associated to the primitive cubic cell.

We also performed an elastic tensor calculation employing a 6×6×6 K-point grid, an energy cutoff of 800eV and a step size of 0.015Å for the central differences. The strain-stress method as implemented on VASP was applied to the primitive unit cell \cite{vasp_stresstrain}. Elastic properties were afterwards extracted using MechElastic script \cite{mechelastic}.

The phonon dispersion and the phonon density of states were also calculated using density functional perturbation theory as implemented in VASP to compute the Hessian matrix, thus, obtaining the interatomic force constants. A supercell of 2x2x2 (regarding the primitive cubic cell) and 3x3x3 K-point grid were used the phonon dispersion was then obtained using the Phonopy code \cite{phonopy}. 

In addition, atomic orbital computations at the HSE06-6-31G(d,p) level \cite{hse} were performed with Gaussian09 \cite{g09}. 


The new $\mathrm{C}_{60}$-based carbon clathrate structure was found in DFT investigations of possible intermolecular bonds formed between $\mathrm{C}_{60}$ molecules in the fcc lattice. A new intermolecular bonding was considered, in which bonds are formed between pentagons of neighboring molecules. The new bonding has four intermolecular covalent bonds and is denoted “double 5/5 2+3 cycloaddition” joining two pentagonal rings in an antiparallel fashion, Figures 1a) and S1. Strout and coworkers reported a similar bonding, although it involves hexagons instead of pentagons \cite{strout1993dim}. A new polymeric structure is then constructed with each molecule adopting a standard molecular orientation and bonded, through the double 5/5 2+3 cycloadditions, to its twelve near neighbors (NN) in the fcc lattice, as illustrated in Figure 1b).

\begin{figure}[h]
	\centering
	\includegraphics[scale=0.27]{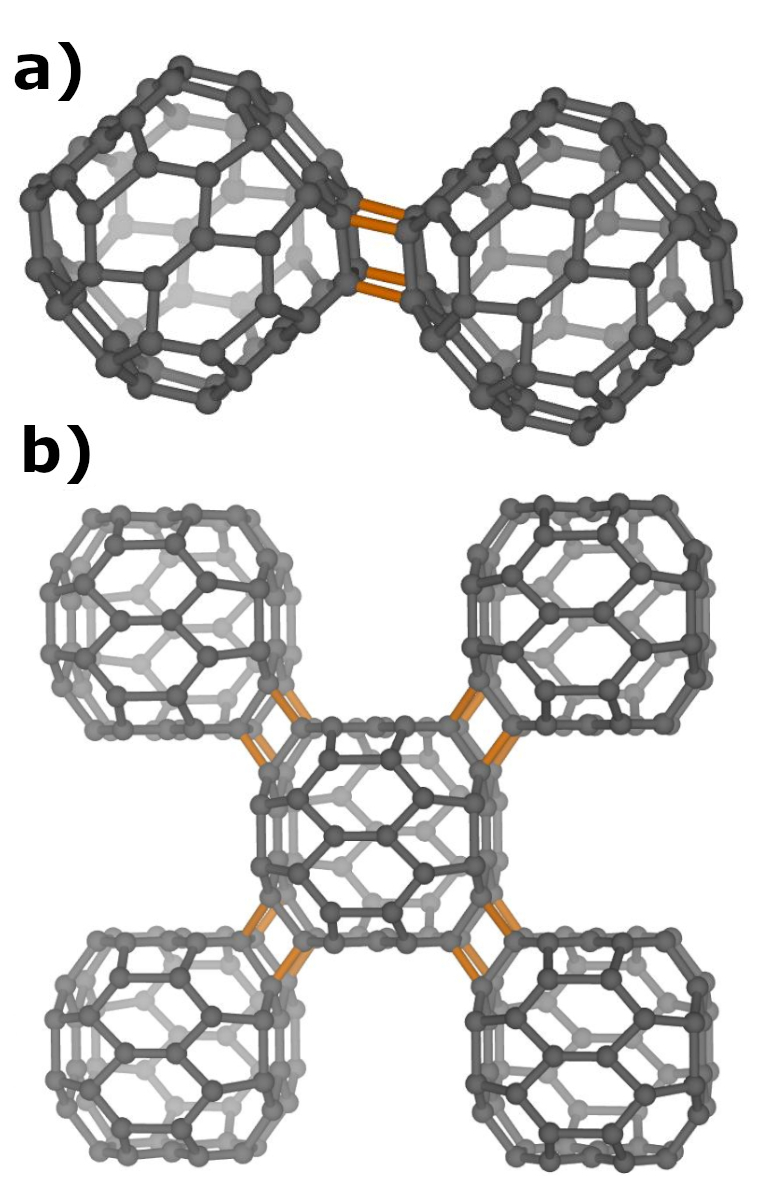}
	\caption{a) Double 5/5 2+3 cycloaddition bond established between neighboring molecules. b) Molecules in the fcc (100) plane bonded through double 5/5 2+3 cycloadditions. Intermolecular bonds are drawn in orange.}
	\label{fig1}
\end{figure}

An important consequence of having each molecule bonded to its neighbors via the double 5/5 2+3 cycloadditions, is that the resulting fcc polymeric structure has its octahedral interstitial sites enclosed by truncated icosahedron cages (identical to the original $\mathrm{C}_{60}$ molecular cages) and its tetrahedral interstitial sites enclosed by distorted truncated octahedron cages. These truncated octahedron cages, with 24 atoms arranged in 8 hexagons and 6 rhombi, are a distorted version of the sodalite-type cages displayed by numerous clathrates \cite{strobel2020,Zhu2020,tao2015,li2016_base,liu2017cla}. The original $\mathrm{C}_{60}$ molecular cages are strongly deformed and their double hexagonal rings, lying perpendicular to the cubic axes, are now perfectly planar. The $\mathrm{C}_{60}$ cages formed on the octahedral sites are exactly equal to the original $\mathrm{C}_{60}$ molecular cages. Hence, as it is illustrated in Figure 2a), the resulting $\mathrm{F}{m}\bar{3}$ fcc structure can be described by a reduced $\mathrm{P}{m}\bar{3}$ simple cubic structure, with half of the lattice constant. The $\mathrm{C}_{60}$ cages share with each other the planar double hexagonal rings, while they share hexagons with the $\mathrm{C}_{24}$ sodalite-like cages, thus, forming a carbon clathrate structure. Figure 2, b) and c), illustrates the cages exhibited by this clathrate structure, emphasizing the different origins of the two $\mathrm{C}_{60}$ cages, one corresponding to the original molecules and the other, centered at the octahedral sites, resulting from the polymeric bonding.

\begin{figure*}[t]
	\centering
	\includegraphics[width=\textwidth]{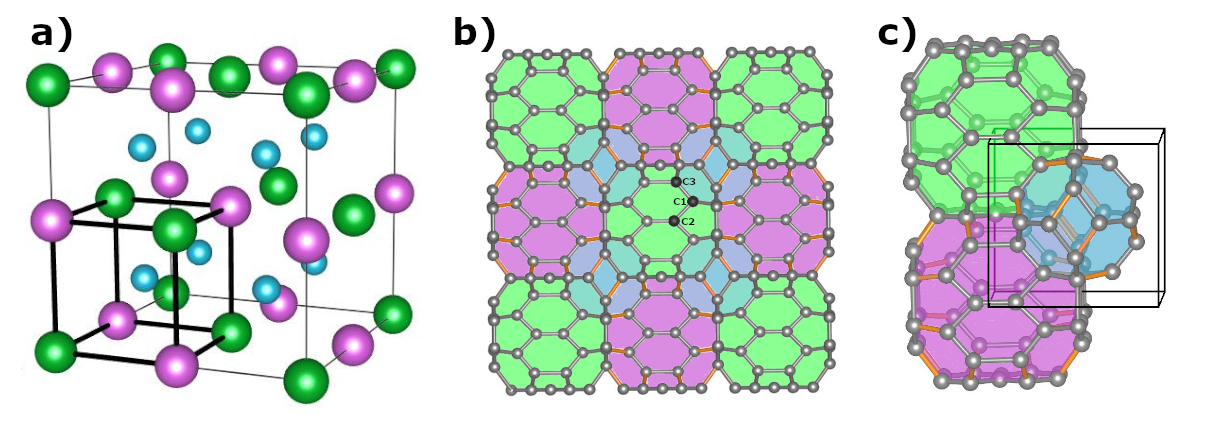}
	\caption{a) Face centered cubic lattice with green spheres as lattice points, lilac spheres as octahedral interstitial sites and blue spheres as tetrahedral interstitial sites. A simple cubic lattice can be drawn when the fcc lattice points and octahedral fcc sites are considered as identical lattice points. b) The $\mathrm{C}_{60}$-based clathrate structure viewed along the [100] direction. $\mathrm{C}_{60}$ cages enclosing fcc lattice points are green, $\mathrm{C}_{60}$ cages enclosing octahedral fcc interstitial sites are lilac, and $\mathrm{C}_{24}$ sodalite-like cages enclosing tetrahedral fcc interstitial sites are blue. c) $\mathrm{C}_{60}$-based clathrate structure described in the simple cubic cell, viewed along [100] direction, with only two $\mathrm{C}_{60}$ cages shown and, at the center of the cell, a $\mathrm{C}_{24}$ sodalite-like cage (the colors have a correspondence as used before). Intermolecular bonds are drawn in orange.}
	\label{fig2}
\end{figure*}

The structural optimization carried out on the fcc lattice, at zero pressure and without symmetry constraints, gives a lattice constant of 12.42\AA, the standard molecular orientation remains unaltered during the optimization. A structural optimization conducted in the simple cubic cell gives exactly the same relaxed structure with a lattice constant of 6.21\AA, half of the relaxed fcc cell. The optimized atomic coordinates of the asymmetric cell, together with their Wyckoff positions, are given in table~\ref{tab:table1} for the simple cubic structure. The C2 atoms are $\mathrm{sp}^{2}$-hybridized, whereas both C1 and C3 atoms are $\mathrm{sp}^{3}$-hybridized, and they are depicted in Figure 2 b). Similar values were obtained from HSE06-6-31G(d,p) level calculations, table S1.

\begin{table}[h]
\caption{\label{tab:table1}%
Optimized atomic positions of the $\mathrm{C}_{60}$-based clathrate structure in the $\mathrm{P}{m}\bar{3}$ simple cubic cell.
}
\begin{ruledtabular}
\begin{tabular}{c c c c c} 
\colrule
	atom & x & y & z  & Wyckoff\\
 				 &  &  &  &Position \\
 		\hline
 		C1  &  0.50000 &  0.29165 & -0.17136 & 12k    \\
 		C2  & -0.12132 &  0.00000 &  0.50000 & 6f   \\
 		C3  &  0.50000 &  0.14222 &  0.34897 & 12k  \\
\end{tabular}
\end{ruledtabular}
\end{table}

The curve of the total energy per atom E(V) for the new $\mathrm{C}_{60}$-based carbon clathrate as a function of volume is given in figure S2. The dynamic stability of the new clathrate structure was also confirmed, since no negative frequencies were found (the phonon dispersion and phonon density of states are given in Figure S3). The total energy calculated for the equilibrium clathrate structure is -7.93eV/atom, significantly larger than the -8.85eV/atom calculated for the $\mathrm{C}_{60}$ molecule. This new structure is also less stable than other 3D $\mathrm{C}_{60}$ polymer structures \cite{laran2018}. 

The electronic band structure and the density of states (DOS) are shown in Figure 3. The carbon clathrate is a narrow-gap semiconductor with an indirect band gap of 0.68eV between the bottom of the conduction band at the M point and the top of the valence band at the X point. This value was also computed at the HSE06-6-31G(d,p) level yielding 1.33eV.

\begin{figure}[h]
	\centering
	\includegraphics[scale=0.22]{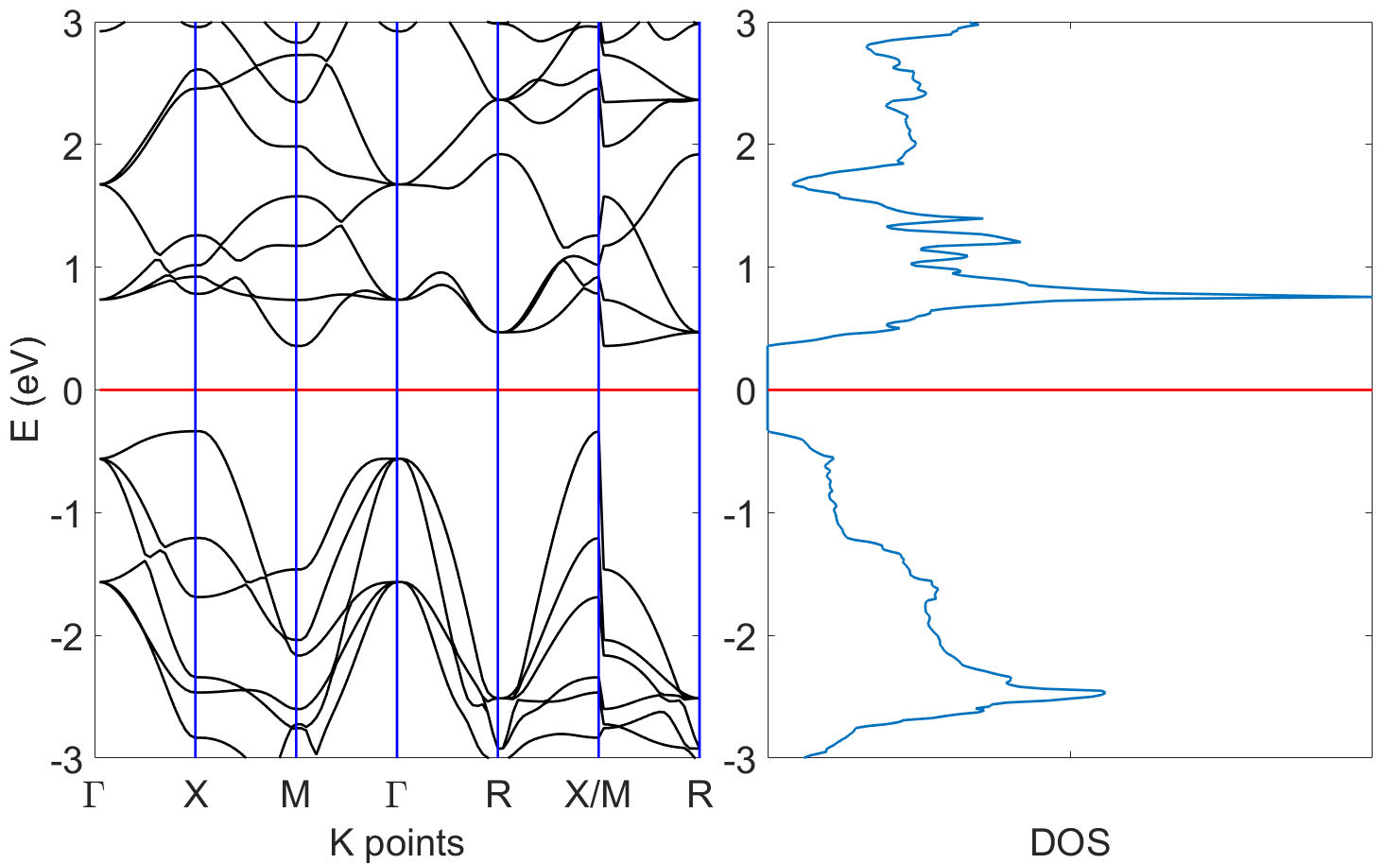}
	\caption{Electronic density of states (right panel) and electronic band structure (left panel) of $\mathrm{C}_{60}$-based clathrate. Red lines indicate the Fermi level.}
	\label{figdos}
\end{figure}

The high proportion of $\mathrm{sp}^{3}$ carbons present in this structure (80\%) should be the reason for the observed semiconducting behavior. The three bands located just above the Fermi level and degenerate at the $\Gamma$ point, are nearly flat and originate a huge peak in DOS at around 0.76eV. Doping, for instance with trivalent metallic atoms inside the $\mathrm{C}_{60}$ cages, would push the Fermi level to this peak, inducing metallic behavior and possibly superconductivity \cite{damien2010}. The corresponding electronic density shown in figure S4 indicates that these three bands are nearly pure antibonding $\pi^*$ orbitals of the $\mathrm{sp}^{2}$ carbons (C2 atoms in Figure 2b)).

The mechanical stability of the $\mathrm{C}_{60}$-based clathrate structure was evaluated by computing the elastic tensor and checking the elastic stability criteria, which are the following for the cubic symmetry: $C_{11}-C_{12}>0,$\space $ C_{11}+2C_{12}>0,$\space $ C_{44}>0$ \cite{elasticcretiria}. The computed values for these elastic constants are $C_{11}$=446GPa, $C_{12}$=179GPa and $C_{44}$=207GPa and therefore all the conditions of stability are satisfied. We used these values to compute the elastic properties, in the framework of Voigt- Reuss-Hill approximation, obtaining $B_0=268$GPa, G=174GPa, E=429GPa and $\nu=0.23$, for the bulk, shear and Young moduli, and the Poisson ratio, respectively \cite{Hill_1952}.

We have also estimated the clathrate hardness using the macroscopic hardness model developed by Chen and coworkers, as implemented in MechElastic code \cite{chenhardness,mechelastic}. A value of 21.6GPa was found, in agreement with the observed moderate elastic constants \cite{li2016_base,soldatov2020}. The $\mathrm{C}_{60}$-based carbon clathrate with bigger cages and 20\% of $\mathrm{sp}^{2}$ carbons has a lower mechanical performance than the reported theoretical carbon clathrates, having smaller cages and being fully $\mathrm{sp}^{3}$-bonded.

High-pressure and high-temperature (HPHT) are known to induce ordered polymerization of $\mathrm{C}_{60}$ molecules. Indeed, with increasing pressure synthesis, one-dimensional and two-dimensional $\mathrm{C}_{60}$ polymers have been prepared; inducing 3D polymerization require higher pressures, above 10GPa. The clathrate structure reported here is, therefore, likely to be prepared by subjecting fullerite $\mathrm{C}_{60}$ to HPHT conditions above 10GPa. Its density, 2.5$\mathrm{g/cm}^3$, is close to densities reported for the 3D $\mathrm{C}_{60}$ polymers already synthesized. However, in the new $\mathrm{C}_{60}$-based carbon clathrate eighty percent of its atoms are $\mathrm{sp}^{3}$-hybridized, a proportion higher than it was experimentally reported for 3D polymers. Yamanaka and coworkers found two 3D polymerized phases, with Immm and $\mathrm{R}\bar{3}$m  symmetries, both obtained at 15GPa and 823 K, containing 49\% and 45\% of $\mathrm{sp}^{3}$ carbons, respectively \cite{yamaka_sp3}. On the other hand Lyapin and coworkers reported a cubic 3D $\mathrm{C}_{60}$ polymer obtained at 12.5GPa with 60\% of its atoms in $\mathrm{sp}^{3}$ hybridization state [25]. Recently, Zhang and coworkers subjected fullerite $\mathrm{C}_{60}$ to 25GPa to synthesize three amorphous phases with an estimated ratio of $\mathrm{sp}^{3}$ carbons between 43\% and 72\% \cite{soldatov2020}.

The elastic properties of the new clathrate structure, in particular the calculated bulk modulus of 268GPa, are close to those reported for 3D $\mathrm{C}_{60}$ polymers. Lyapin and coworkers \cite{lyapin2019} reported a bulk modulus of 280GPa, although the published data has a significant uncertainty. Mezouar and coworkers \cite{mezouarprb2003} also reported a 3D polymer with a bulk modulus of 288GPa.

It is then plausible that $\mathrm{C}_{60}$ 3D unresolved polymer phases might consist or include carbon clathrates related to the one reported here. This possibility, if confirmed, would highlight the relevance of high-pressure in inducing or stabilizing carbon clathrate structures.

In summary, a novel $\mathrm{C}_{60}$-based carbon clathrate was predicted through DFT investigations of $\mathrm{C}_{60}$ polymerization in the solid state. To the best of our knowledge, this is the first proposed carbon clathrate structure with $\mathrm{C}_{60}$ molecular building blocks and, since it results from $\mathrm{C}_{60}$ polymerization, it is quite likely to be prepared through high pressure high temperature treatment of fullerite $\mathrm{C}_{60}$.

The high number of $\mathrm{sp}^{3}$ carbons in this phase leads to a semiconducting material with moderate elastic properties. Yet, appropriate doping, for instance, using endohedral metallofullerenes, may render this phase metallic and superconducting with properties potentially similar barium silicon clathrate and the new metal hydrides, which display room-temperature superconductivity at megabar pressures.

\begin{acknowledgments}
This work was supported by the projects POCI-01-0145-FEDER-031326 and IF/00894/2015 financed by the Portuguese Foundation for Science and Technology (FCT) and co-financed by FEDER. CICECO-Aveiro Institute of Materials, UIDB/50011/2020 \& UIDP/50011/2020, financed by national funds through the FCT/MEC and when appropriate co-financed by FEDER under the PT2020 Partnership Agreement. J. Laranjeira acknowledges a PhD grant from FCT (SFRH/BD/139327/2018).
\end{acknowledgments}





\bibliography{apssamp}

\begin{thebibliography}{38}%
\makeatletter
\providecommand \@ifxundefined [1]{%
 \@ifx{#1\undefined}
}%
\providecommand \@ifnum [1]{%
 \ifnum #1\expandafter \@firstoftwo
 \else \expandafter \@secondoftwo
 \fi
}%
\providecommand \@ifx [1]{%
 \ifx #1\expandafter \@firstoftwo
 \else \expandafter \@secondoftwo
 \fi
}%
\providecommand \natexlab [1]{#1}%
\providecommand \enquote  [1]{``#1''}%
\providecommand \bibnamefont  [1]{#1}%
\providecommand \bibfnamefont [1]{#1}%
\providecommand \citenamefont [1]{#1}%
\providecommand \href@noop [0]{\@secondoftwo}%
\providecommand \href [0]{\begingroup \@sanitize@url \@href}%
\providecommand \@href[1]{\@@startlink{#1}\@@href}%
\providecommand \@@href[1]{\endgroup#1\@@endlink}%
\providecommand \@sanitize@url [0]{\catcode `\\12\catcode `\$12\catcode
  `\&12\catcode `\#12\catcode `\^12\catcode `\_12\catcode `\%12\relax}%
\providecommand \@@startlink[1]{}%
\providecommand \@@endlink[0]{}%
\providecommand \url  [0]{\begingroup\@sanitize@url \@url }%
\providecommand \@url [1]{\endgroup\@href {#1}{\urlprefix }}%
\providecommand \urlprefix  [0]{URL }%
\providecommand \Eprint [0]{\href }%
\providecommand \doibase [0]{https://doi.org/}%
\providecommand \selectlanguage [0]{\@gobble}%
\providecommand \bibinfo  [0]{\@secondoftwo}%
\providecommand \bibfield  [0]{\@secondoftwo}%
\providecommand \translation [1]{[#1]}%
\providecommand \BibitemOpen [0]{}%
\providecommand \bibitemStop [0]{}%
\providecommand \bibitemNoStop [0]{.\EOS\space}%
\providecommand \EOS [0]{\spacefactor3000\relax}%
\providecommand \BibitemShut  [1]{\csname bibitem#1\endcsname}%
\let\auto@bib@innerbib\@empty
\bibitem [{\citenamefont {Hassanpouryouzband}\ \emph
  {et~al.}(2020)\citenamefont {Hassanpouryouzband}, \citenamefont {Joonaki},
  \citenamefont {Farahani}, \citenamefont {Takeya}, \citenamefont {Ruppel},
  \citenamefont {Yang}, \citenamefont {English}, \citenamefont {Schicks},
  \citenamefont {Edlmann}, \citenamefont {Mehrabian}, \citenamefont {Aman},\
  and\ \citenamefont {Tohidi}}]{gashydrysdescsr}%
  \BibitemOpen
  \bibfield  {author} {\bibinfo {author} {\bibfnamefont {A.}~\bibnamefont
  {Hassanpouryouzband}}, \bibinfo {author} {\bibfnamefont {E.}~\bibnamefont
  {Joonaki}}, \bibinfo {author} {\bibfnamefont {M.}~\bibnamefont {Farahani}},
  \bibinfo {author} {\bibfnamefont {S.}~\bibnamefont {Takeya}}, \bibinfo
  {author} {\bibfnamefont {C.}~\bibnamefont {Ruppel}}, \bibinfo {author}
  {\bibfnamefont {J.}~\bibnamefont {Yang}}, \bibinfo {author} {\bibfnamefont
  {N.}~\bibnamefont {English}}, \bibinfo {author} {\bibfnamefont
  {J.}~\bibnamefont {Schicks}}, \bibinfo {author} {\bibfnamefont
  {K.}~\bibnamefont {Edlmann}}, \bibinfo {author} {\bibfnamefont
  {H.}~\bibnamefont {Mehrabian}}, \bibinfo {author} {\bibfnamefont
  {Z.}~\bibnamefont {Aman}},\ and\ \bibinfo {author} {\bibfnamefont
  {B.}~\bibnamefont {Tohidi}},\ }\bibfield  {title} {\bibinfo {title} {Gas
  hydrates in sustainable chemistry},\ }\href
  {https://doi.org/http://dx.doi.org/10.1039/C8CS00989A} {\bibfield  {journal}
  {\bibinfo  {journal} {Chem. Soc. Rev.}\ }\textbf {\bibinfo {volume} {49}},\
  \bibinfo {pages} {5225} (\bibinfo {year} {2020})}\BibitemShut {NoStop}%
\bibitem [{\citenamefont {Guloy}\ \emph {et~al.}(2006)\citenamefont {Guloy},
  \citenamefont {Ramlau}, \citenamefont {Tang}, \citenamefont {Schnelle},
  \citenamefont {Baitinger},\ and\ \citenamefont {Grin}}]{geclatrateexp}%
  \BibitemOpen
  \bibfield  {author} {\bibinfo {author} {\bibfnamefont {A.}~\bibnamefont
  {Guloy}}, \bibinfo {author} {\bibfnamefont {R.}~\bibnamefont {Ramlau}},
  \bibinfo {author} {\bibfnamefont {Z.}~\bibnamefont {Tang}}, \bibinfo {author}
  {\bibfnamefont {W.}~\bibnamefont {Schnelle}}, \bibinfo {author}
  {\bibfnamefont {M.}~\bibnamefont {Baitinger}},\ and\ \bibinfo {author}
  {\bibfnamefont {Y.}~\bibnamefont {Grin}},\ }\bibfield  {title} {\bibinfo
  {title} {A guest-free germanium clathrate},\ }\href
  {https://doi.org/https://doi.org/10.1038/nature05145} {\bibfield  {journal}
  {\bibinfo  {journal} {Nature}\ }\textbf {\bibinfo {volume} {443}},\ \bibinfo
  {pages} {320} (\bibinfo {year} {2006})}\BibitemShut {NoStop}%
\bibitem [{\citenamefont {Yamanaka}(2010)}]{yamanakaclatrato}%
  \BibitemOpen
  \bibfield  {author} {\bibinfo {author} {\bibfnamefont {S.}~\bibnamefont
  {Yamanaka}},\ }\bibfield  {title} {\bibinfo {title} {Silicon clathrates and
  carbon analogs: high pressure synthesis{,} structure{,} and
  superconductivity},\ }\href {https://doi.org/10.1039/B918480E} {\bibfield
  {journal} {\bibinfo  {journal} {Dalton Trans.}\ }\textbf {\bibinfo {volume}
  {39}},\ \bibinfo {pages} {1901} (\bibinfo {year} {2010})}\BibitemShut
  {NoStop}%
\bibitem [{\citenamefont {Sun}\ \emph {et~al.}(2019)\citenamefont {Sun},
  \citenamefont {Lv}, \citenamefont {Xie}, \citenamefont {Liu},\ and\
  \citenamefont {Ma}}]{sun_hyd}%
  \BibitemOpen
  \bibfield  {author} {\bibinfo {author} {\bibfnamefont {Y.}~\bibnamefont
  {Sun}}, \bibinfo {author} {\bibfnamefont {J.}~\bibnamefont {Lv}}, \bibinfo
  {author} {\bibfnamefont {Y.}~\bibnamefont {Xie}}, \bibinfo {author}
  {\bibfnamefont {H.}~\bibnamefont {Liu}},\ and\ \bibinfo {author}
  {\bibfnamefont {Y.}~\bibnamefont {Ma}},\ }\bibfield  {title} {\bibinfo
  {title} {Route to a superconducting phase above room temperature in
  electron-doped hydride compounds under high pressure},\ }\href
  {https://doi.org/10.1103/PhysRevLett.123.097001} {\bibfield  {journal}
  {\bibinfo  {journal} {Phys. Rev. Lett.}\ }\textbf {\bibinfo {volume} {123}},\
  \bibinfo {pages} {097001} (\bibinfo {year} {2019})}\BibitemShut {NoStop}%
\bibitem [{\citenamefont {Snider}\ \emph {et~al.}(2020)\citenamefont {Snider},
  \citenamefont {Dasenbrock-Gammon}, \citenamefont {McBride}, \citenamefont
  {Debessai}, \citenamefont {Vindana}, \citenamefont {Vencatasamy},
  \citenamefont {Lawler}, \citenamefont {Salamat},\ and\ \citenamefont
  {Dias}}]{dias_hyd}%
  \BibitemOpen
  \bibfield  {author} {\bibinfo {author} {\bibfnamefont {E.}~\bibnamefont
  {Snider}}, \bibinfo {author} {\bibfnamefont {N.}~\bibnamefont
  {Dasenbrock-Gammon}}, \bibinfo {author} {\bibfnamefont {R.}~\bibnamefont
  {McBride}}, \bibinfo {author} {\bibfnamefont {M.}~\bibnamefont {Debessai}},
  \bibinfo {author} {\bibfnamefont {H.}~\bibnamefont {Vindana}}, \bibinfo
  {author} {\bibfnamefont {K.}~\bibnamefont {Vencatasamy}}, \bibinfo {author}
  {\bibfnamefont {K.}~\bibnamefont {Lawler}}, \bibinfo {author} {\bibfnamefont
  {A.}~\bibnamefont {Salamat}},\ and\ \bibinfo {author} {\bibfnamefont
  {R.}~\bibnamefont {Dias}},\ }\bibfield  {title} {\bibinfo {title}
  {Room-temperature superconductivity in a carbonaceous sulfur hydride},\
  }\href {https://doi.org/10.1038/s41586-020-2801-z} {\bibfield  {journal}
  {\bibinfo  {journal} {Nature}\ }\textbf {\bibinfo {volume} {586}},\ \bibinfo
  {pages} {373} (\bibinfo {year} {2020})}\BibitemShut {NoStop}%
\bibitem [{\citenamefont {Somayazulu}\ \emph {et~al.}(2019)\citenamefont
  {Somayazulu}, \citenamefont {Ahart}, \citenamefont {Mishra}, \citenamefont
  {Geballe}, \citenamefont {Baldini}, \citenamefont {Meng}, \citenamefont
  {Struzhkin},\ and\ \citenamefont {Hemley}}]{hemley_hyd}%
  \BibitemOpen
  \bibfield  {author} {\bibinfo {author} {\bibfnamefont {M.}~\bibnamefont
  {Somayazulu}}, \bibinfo {author} {\bibfnamefont {M.}~\bibnamefont {Ahart}},
  \bibinfo {author} {\bibfnamefont {A.}~\bibnamefont {Mishra}}, \bibinfo
  {author} {\bibfnamefont {Z.}~\bibnamefont {Geballe}}, \bibinfo {author}
  {\bibfnamefont {M.}~\bibnamefont {Baldini}}, \bibinfo {author} {\bibfnamefont
  {Y.}~\bibnamefont {Meng}}, \bibinfo {author} {\bibfnamefont {V.}~\bibnamefont
  {Struzhkin}},\ and\ \bibinfo {author} {\bibfnamefont {R.}~\bibnamefont
  {Hemley}},\ }\bibfield  {title} {\bibinfo {title} {Evidence for
  superconductivity above 260 $\mathrm{K}$ in lanthanum superhydride at megabar
  pressures},\ }\href {https://doi.org/10.1103/PhysRevLett.122.027001}
  {\bibfield  {journal} {\bibinfo  {journal} {Phys. Rev. Lett.}\ }\textbf
  {\bibinfo {volume} {122}},\ \bibinfo {pages} {027001} (\bibinfo {year}
  {2019})}\BibitemShut {NoStop}%
\bibitem [{\citenamefont {Drozdov}\ \emph {et~al.}(2019)\citenamefont
  {Drozdov}, \citenamefont {Kong}, \citenamefont {Minkov}, \citenamefont
  {Besedin}, \citenamefont {Kuzovnikov}, \citenamefont {Mozaffari},
  \citenamefont {Balicas}, \citenamefont {Balakirev}, \citenamefont {Graf},
  \citenamefont {Prakapenka}, \citenamefont {Greenberg}, \citenamefont
  {Knyazev}, \citenamefont {Tkacz},\ and\ \citenamefont
  {Eremets}}]{eremets_hyd}%
  \BibitemOpen
  \bibfield  {author} {\bibinfo {author} {\bibfnamefont {A.}~\bibnamefont
  {Drozdov}}, \bibinfo {author} {\bibfnamefont {P.}~\bibnamefont {Kong}},
  \bibinfo {author} {\bibfnamefont {V.}~\bibnamefont {Minkov}}, \bibinfo
  {author} {\bibfnamefont {S.}~\bibnamefont {Besedin}}, \bibinfo {author}
  {\bibfnamefont {M.}~\bibnamefont {Kuzovnikov}}, \bibinfo {author}
  {\bibfnamefont {S.}~\bibnamefont {Mozaffari}}, \bibinfo {author}
  {\bibfnamefont {L.}~\bibnamefont {Balicas}}, \bibinfo {author} {\bibfnamefont
  {F.}~\bibnamefont {Balakirev}}, \bibinfo {author} {\bibfnamefont
  {D.}~\bibnamefont {Graf}}, \bibinfo {author} {\bibfnamefont {V.}~\bibnamefont
  {Prakapenka}}, \bibinfo {author} {\bibfnamefont {E.}~\bibnamefont
  {Greenberg}}, \bibinfo {author} {\bibfnamefont {D.}~\bibnamefont {Knyazev}},
  \bibinfo {author} {\bibfnamefont {M.}~\bibnamefont {Tkacz}},\ and\ \bibinfo
  {author} {\bibfnamefont {M.}~\bibnamefont {Eremets}},\ }\bibfield  {title}
  {\bibinfo {title} {Superconductivity at 250 $\mathrm{K}$ in lanthanum hydride
  under high pressures},\ }\href {https://doi.org/10.1038/s41586-019-1201-8}
  {\bibfield  {journal} {\bibinfo  {journal} {Nature}\ }\textbf {\bibinfo
  {volume} {569}},\ \bibinfo {pages} {528} (\bibinfo {year}
  {2019})}\BibitemShut {NoStop}%
\bibitem [{\citenamefont {Pickard}\ \emph {et~al.}(2020)\citenamefont
  {Pickard}, \citenamefont {Errea},\ and\ \citenamefont
  {Eremets}}]{pickard_hydrides}%
  \BibitemOpen
  \bibfield  {author} {\bibinfo {author} {\bibfnamefont {C.}~\bibnamefont
  {Pickard}}, \bibinfo {author} {\bibfnamefont {I.}~\bibnamefont {Errea}},\
  and\ \bibinfo {author} {\bibfnamefont {M.}~\bibnamefont {Eremets}},\
  }\bibfield  {title} {\bibinfo {title} {Superconducting hydrides under
  pressure},\ }\href {https://doi.org/10.1146/annurev-conmatphys-031218-013413}
  {\bibfield  {journal} {\bibinfo  {journal} {Rev. Condens. Matter Phys.}\
  }\textbf {\bibinfo {volume} {11}},\ \bibinfo {pages} {57} (\bibinfo {year}
  {2020})}\BibitemShut {NoStop}%
\bibitem [{\citenamefont {Strobel}\ \emph {et~al.}(2020)\citenamefont
  {Strobel}, \citenamefont {Zhu}, \citenamefont {Guńka}, \citenamefont
  {Borstad},\ and\ \citenamefont {Guerette}}]{strobel2020}%
  \BibitemOpen
  \bibfield  {author} {\bibinfo {author} {\bibfnamefont {T.}~\bibnamefont
  {Strobel}}, \bibinfo {author} {\bibfnamefont {L.}~\bibnamefont {Zhu}},
  \bibinfo {author} {\bibfnamefont {P.}~\bibnamefont {Guńka}}, \bibinfo
  {author} {\bibfnamefont {G.}~\bibnamefont {Borstad}},\ and\ \bibinfo {author}
  {\bibfnamefont {M.}~\bibnamefont {Guerette}},\ }\bibfield  {title} {\bibinfo
  {title} {{A Lanthanum-Filled Carbon–Boron Clathrate}},\ }\href
  {https://doi.org/https://doi.org/10.1002/anie.202012821} {\bibfield
  {journal} {\bibinfo  {journal} {Angew. Chem. Int. Ed.}\ }\textbf {\bibinfo
  {volume} {60}},\ \bibinfo {pages} {2877} (\bibinfo {year}
  {2020})}\BibitemShut {NoStop}%
\bibitem [{\citenamefont {Zhu}\ \emph {et~al.}(2020)\citenamefont {Zhu},
  \citenamefont {Borstad}, \citenamefont {Liu}, \citenamefont {Gu{\'n}ka},
  \citenamefont {Guerette}, \citenamefont {Dolyniuk}, \citenamefont {Meng},
  \citenamefont {Greenberg}, \citenamefont {Prakapenka}, \citenamefont
  {Chaloux}, \citenamefont {Epshteyn}, \citenamefont {Cohen},\ and\
  \citenamefont {Strobel}}]{Zhu2020}%
  \BibitemOpen
  \bibfield  {author} {\bibinfo {author} {\bibfnamefont {L.}~\bibnamefont
  {Zhu}}, \bibinfo {author} {\bibfnamefont {G.}~\bibnamefont {Borstad}},
  \bibinfo {author} {\bibfnamefont {H.}~\bibnamefont {Liu}}, \bibinfo {author}
  {\bibfnamefont {P.}~\bibnamefont {Gu{\'n}ka}}, \bibinfo {author}
  {\bibfnamefont {M.}~\bibnamefont {Guerette}}, \bibinfo {author}
  {\bibfnamefont {J.-A.}\ \bibnamefont {Dolyniuk}}, \bibinfo {author}
  {\bibfnamefont {Y.}~\bibnamefont {Meng}}, \bibinfo {author} {\bibfnamefont
  {E.}~\bibnamefont {Greenberg}}, \bibinfo {author} {\bibfnamefont
  {V.}~\bibnamefont {Prakapenka}}, \bibinfo {author} {\bibfnamefont
  {B.}~\bibnamefont {Chaloux}}, \bibinfo {author} {\bibfnamefont
  {A.}~\bibnamefont {Epshteyn}}, \bibinfo {author} {\bibfnamefont
  {R.}~\bibnamefont {Cohen}},\ and\ \bibinfo {author} {\bibfnamefont
  {T.}~\bibnamefont {Strobel}},\ }\bibfield  {title} {\bibinfo {title}
  {Carbon-boron clathrates as a new class of $\mathrm{sp}^{3}$-bonded framework
  materials},\ }\href@noop {} {\bibfield  {journal} {\bibinfo  {journal} {Sci.
  Adv.}\ }\textbf {\bibinfo {volume} {6}} (\bibinfo {year} {2020})}\BibitemShut
  {NoStop}%
\bibitem [{\citenamefont {Zeng}\ \emph {et~al.}(2015)\citenamefont {Zeng},
  \citenamefont {Hoffmann}, \citenamefont {Nesper}, \citenamefont {Ashcroft},
  \citenamefont {Strobel},\ and\ \citenamefont {Proserpio}}]{tao2015}%
  \BibitemOpen
  \bibfield  {author} {\bibinfo {author} {\bibfnamefont {T.}~\bibnamefont
  {Zeng}}, \bibinfo {author} {\bibfnamefont {R.}~\bibnamefont {Hoffmann}},
  \bibinfo {author} {\bibfnamefont {R.}~\bibnamefont {Nesper}}, \bibinfo
  {author} {\bibfnamefont {N.}~\bibnamefont {Ashcroft}}, \bibinfo {author}
  {\bibfnamefont {T.}~\bibnamefont {Strobel}},\ and\ \bibinfo {author}
  {\bibfnamefont {D.}~\bibnamefont {Proserpio}},\ }\bibfield  {title} {\bibinfo
  {title} {Li-filled, $\mathrm{B}$-substituted carbon clathrates},\ }\href
  {https://doi.org/10.1021/jacs.5b07883} {\bibfield  {journal} {\bibinfo
  {journal} {J. Am. Chem. Soc.}\ }\textbf {\bibinfo {volume} {137}},\ \bibinfo
  {pages} {12639} (\bibinfo {year} {2015})}\BibitemShut {NoStop}%
\bibitem [{\citenamefont {Liu}\ \emph {et~al.}(2017)\citenamefont {Liu},
  \citenamefont {Jiang}, \citenamefont {Huang}, \citenamefont {Zhou},\ and\
  \citenamefont {Zhao}}]{liu2017cla}%
  \BibitemOpen
  \bibfield  {author} {\bibinfo {author} {\bibfnamefont {Y.}~\bibnamefont
  {Liu}}, \bibinfo {author} {\bibfnamefont {X.}~\bibnamefont {Jiang}}, \bibinfo
  {author} {\bibfnamefont {Y.}~\bibnamefont {Huang}}, \bibinfo {author}
  {\bibfnamefont {S.}~\bibnamefont {Zhou}},\ and\ \bibinfo {author}
  {\bibfnamefont {J.}~\bibnamefont {Zhao}},\ }\bibfield  {title} {\bibinfo
  {title} {A new family of multifunctional silicon clathrates: Optoelectronic
  and thermoelectric applications},\ }\href@noop {} {\bibfield  {journal}
  {\bibinfo  {journal} {J. Appl. Phys.}\ }\textbf {\bibinfo {volume} {121}},\
  \bibinfo {pages} {085107} (\bibinfo {year} {2017})}\BibitemShut {NoStop}%
\bibitem [{\citenamefont {Li}\ \emph {et~al.}(2016)\citenamefont {Li},
  \citenamefont {Hu}, \citenamefont {Ma}, \citenamefont {Gao}, \citenamefont
  {Xu}, \citenamefont {He}, \citenamefont {Yu}, \citenamefont {Tian},\ and\
  \citenamefont {Zhao}}]{li2016_base}%
  \BibitemOpen
  \bibfield  {author} {\bibinfo {author} {\bibfnamefont {Z.}~\bibnamefont
  {Li}}, \bibinfo {author} {\bibfnamefont {M.}~\bibnamefont {Hu}}, \bibinfo
  {author} {\bibfnamefont {M.}~\bibnamefont {Ma}}, \bibinfo {author}
  {\bibfnamefont {Y.}~\bibnamefont {Gao}}, \bibinfo {author} {\bibfnamefont
  {B.}~\bibnamefont {Xu}}, \bibinfo {author} {\bibfnamefont {J.}~\bibnamefont
  {He}}, \bibinfo {author} {\bibfnamefont {D.}~\bibnamefont {Yu}}, \bibinfo
  {author} {\bibfnamefont {Y.}~\bibnamefont {Tian}},\ and\ \bibinfo {author}
  {\bibfnamefont {Z.}~\bibnamefont {Zhao}},\ }\bibfield  {title} {\bibinfo
  {title} {Superhard superstrong carbon clathrate},\ }\href
  {https://doi.org/https://doi.org/10.1016/j.carbon.2016.04.038} {\bibfield
  {journal} {\bibinfo  {journal} {Carbon}\ }\textbf {\bibinfo {volume} {105}},\
  \bibinfo {pages} {151} (\bibinfo {year} {2016})}\BibitemShut {NoStop}%
\bibitem [{\citenamefont {Conn\'etable}\ \emph {et~al.}(2003)\citenamefont
  {Conn\'etable}, \citenamefont {Timoshevskii}, \citenamefont {Masenelli},
  \citenamefont {Beille}, \citenamefont {Marcus}, \citenamefont {Barbara},
  \citenamefont {Saitta}, \citenamefont {Rignanese}, \citenamefont {M\'elinon},
  \citenamefont {Yamanaka},\ and\ \citenamefont {Blase}}]{blasesupercond}%
  \BibitemOpen
  \bibfield  {author} {\bibinfo {author} {\bibfnamefont {D.}~\bibnamefont
  {Conn\'etable}}, \bibinfo {author} {\bibfnamefont {V.}~\bibnamefont
  {Timoshevskii}}, \bibinfo {author} {\bibfnamefont {B.}~\bibnamefont
  {Masenelli}}, \bibinfo {author} {\bibfnamefont {J.}~\bibnamefont {Beille}},
  \bibinfo {author} {\bibfnamefont {J.}~\bibnamefont {Marcus}}, \bibinfo
  {author} {\bibfnamefont {B.}~\bibnamefont {Barbara}}, \bibinfo {author}
  {\bibfnamefont {A.}~\bibnamefont {Saitta}}, \bibinfo {author} {\bibfnamefont
  {G.-M.}\ \bibnamefont {Rignanese}}, \bibinfo {author} {\bibfnamefont
  {P.}~\bibnamefont {M\'elinon}}, \bibinfo {author} {\bibfnamefont
  {S.}~\bibnamefont {Yamanaka}},\ and\ \bibinfo {author} {\bibfnamefont
  {X.}~\bibnamefont {Blase}},\ }\bibfield  {title} {\bibinfo {title}
  {Superconductivity in doped $s{p}^{3}$ semiconductors: The case of the
  clathrates},\ }\href {https://doi.org/10.1103/PhysRevLett.91.247001}
  {\bibfield  {journal} {\bibinfo  {journal} {Phys. Rev. Lett.}\ }\textbf
  {\bibinfo {volume} {91}},\ \bibinfo {pages} {247001} (\bibinfo {year}
  {2003})}\BibitemShut {NoStop}%
\bibitem [{\citenamefont {Blase}\ \emph {et~al.}(2004)\citenamefont {Blase},
  \citenamefont {Gillet}, \citenamefont {San~Miguel},\ and\ \citenamefont
  {M\'elinon}}]{blase2004}%
  \BibitemOpen
  \bibfield  {author} {\bibinfo {author} {\bibfnamefont {X.}~\bibnamefont
  {Blase}}, \bibinfo {author} {\bibfnamefont {P.}~\bibnamefont {Gillet}},
  \bibinfo {author} {\bibfnamefont {A.}~\bibnamefont {San~Miguel}},\ and\
  \bibinfo {author} {\bibfnamefont {P.}~\bibnamefont {M\'elinon}},\ }\bibfield
  {title} {\bibinfo {title} {Exceptional ideal strength of carbon clathrates},\
  }\href {https://doi.org/10.1103/PhysRevLett.92.215505} {\bibfield  {journal}
  {\bibinfo  {journal} {Phys. Rev. Lett.}\ }\textbf {\bibinfo {volume} {92}},\
  \bibinfo {pages} {215505} (\bibinfo {year} {2004})}\BibitemShut {NoStop}%
\bibitem [{\citenamefont {Conn\'etable}(2010)}]{damien2010}%
  \BibitemOpen
  \bibfield  {author} {\bibinfo {author} {\bibfnamefont {D.}~\bibnamefont
  {Conn\'etable}},\ }\bibfield  {title} {\bibinfo {title} {First-principles
  calculations of carbon clathrates: Comparison to silicon and germanium
  clathrates},\ }\href {https://doi.org/10.1103/PhysRevB.82.075209} {\bibfield
  {journal} {\bibinfo  {journal} {Phys. Rev. B}\ }\textbf {\bibinfo {volume}
  {82}},\ \bibinfo {pages} {075209} (\bibinfo {year} {2010})}\BibitemShut
  {NoStop}%
\bibitem [{\citenamefont {Machon}\ \emph {et~al.}(2018)\citenamefont {Machon},
  \citenamefont {Pischedda}, \citenamefont {Le~Floch},\ and\ \citenamefont
  {San-Miguel}}]{sanmiguelc60clatratos}%
  \BibitemOpen
  \bibfield  {author} {\bibinfo {author} {\bibfnamefont {D.}~\bibnamefont
  {Machon}}, \bibinfo {author} {\bibfnamefont {V.}~\bibnamefont {Pischedda}},
  \bibinfo {author} {\bibfnamefont {S.}~\bibnamefont {Le~Floch}},\ and\
  \bibinfo {author} {\bibfnamefont {A.}~\bibnamefont {San-Miguel}},\ }\bibfield
   {title} {\bibinfo {title} {Perspective: High pressure transformations in
  nanomaterials and opportunities in material design},\ }\href
  {https://doi.org/10.1063/1.5045563} {\bibfield  {journal} {\bibinfo
  {journal} {J. Appl. Phys.}\ }\textbf {\bibinfo {volume} {124}},\ \bibinfo
  {pages} {160902} (\bibinfo {year} {2018})}\BibitemShut {NoStop}%
\bibitem [{\citenamefont {Sundqvist}(2021)}]{SUNDQVIST2021}%
  \BibitemOpen
  \bibfield  {author} {\bibinfo {author} {\bibfnamefont {B.}~\bibnamefont
  {Sundqvist}},\ }\bibfield  {title} {\bibinfo {title} {Carbon under
  pressure},\ }\href@noop {} {\bibfield  {journal} {\bibinfo  {journal} {Phys.
  Rep.}\ }\textbf {\bibinfo {volume} {909}},\ \bibinfo {pages} {1} (\bibinfo
  {year} {2021})}\BibitemShut {NoStop}%
\bibitem [{\citenamefont {Wang}\ \emph {et~al.}(2018)\citenamefont {Wang},
  \citenamefont {Nie}, \citenamefont {Weng}, \citenamefont {Kawazoe},\ and\
  \citenamefont {Chen}}]{prlwang2018}%
  \BibitemOpen
  \bibfield  {author} {\bibinfo {author} {\bibfnamefont {J.}~\bibnamefont
  {Wang}}, \bibinfo {author} {\bibfnamefont {S.}~\bibnamefont {Nie}}, \bibinfo
  {author} {\bibfnamefont {H.}~\bibnamefont {Weng}}, \bibinfo {author}
  {\bibfnamefont {Y.}~\bibnamefont {Kawazoe}},\ and\ \bibinfo {author}
  {\bibfnamefont {C.}~\bibnamefont {Chen}},\ }\bibfield  {title} {\bibinfo
  {title} {Topological nodal-net semimetal in a graphene network structure},\
  }\href {https://doi.org/10.1103/PhysRevLett.120.026402} {\bibfield  {journal}
  {\bibinfo  {journal} {Phys. Rev. Lett.}\ }\textbf {\bibinfo {volume} {120}},\
  \bibinfo {pages} {026402} (\bibinfo {year} {2018})}\BibitemShut {NoStop}%
\bibitem [{\citenamefont {Laranjeira}\ \emph {et~al.}(2018)\citenamefont
  {Laranjeira}, \citenamefont {Marques}, \citenamefont {Fortunato},
  \citenamefont {Melle-Franco}, \citenamefont {Strutyński},\ and\
  \citenamefont {Barroso}}]{laran2018}%
  \BibitemOpen
  \bibfield  {author} {\bibinfo {author} {\bibfnamefont {J.}~\bibnamefont
  {Laranjeira}}, \bibinfo {author} {\bibfnamefont {L.}~\bibnamefont {Marques}},
  \bibinfo {author} {\bibfnamefont {N.}~\bibnamefont {Fortunato}}, \bibinfo
  {author} {\bibfnamefont {M.}~\bibnamefont {Melle-Franco}}, \bibinfo {author}
  {\bibfnamefont {K.}~\bibnamefont {Strutyński}},\ and\ \bibinfo {author}
  {\bibfnamefont {M.}~\bibnamefont {Barroso}},\ }\bibfield  {title} {\bibinfo
  {title} {Three-dimensional $\mathrm{C}_{60}$ polymers with ordered
  binary-alloy-type structures},\ }\href
  {https://doi.org/https://doi.org/10.1016/j.carbon.2018.05.070} {\bibfield
  {journal} {\bibinfo  {journal} {Carbon}\ }\textbf {\bibinfo {volume} {137}},\
  \bibinfo {pages} {511} (\bibinfo {year} {2018})}\BibitemShut {NoStop}%
\bibitem [{\citenamefont {Strout}\ \emph {et~al.}(1993)\citenamefont {Strout},
  \citenamefont {Murry}, \citenamefont {Xu}, \citenamefont {Eckhoff},
  \citenamefont {Odom},\ and\ \citenamefont {Scuseria}}]{strout1993dim}%
  \BibitemOpen
  \bibfield  {author} {\bibinfo {author} {\bibfnamefont {D.}~\bibnamefont
  {Strout}}, \bibinfo {author} {\bibfnamefont {R.}~\bibnamefont {Murry}},
  \bibinfo {author} {\bibfnamefont {C.}~\bibnamefont {Xu}}, \bibinfo {author}
  {\bibfnamefont {W.}~\bibnamefont {Eckhoff}}, \bibinfo {author} {\bibfnamefont
  {G.}~\bibnamefont {Odom}},\ and\ \bibinfo {author} {\bibfnamefont
  {G.}~\bibnamefont {Scuseria}},\ }\bibfield  {title} {\bibinfo {title} {A
  theoretical study of buckminsterfullerene reaction products:
  $\mathrm{C}_{60}$+$\mathrm{C}_{60}$},\ }\href
  {https://doi.org/https://doi.org/10.1016/0009-2614(93)85686-I} {\bibfield
  {journal} {\bibinfo  {journal} {Chem. Phys. Lett.}\ }\textbf {\bibinfo
  {volume} {214}},\ \bibinfo {pages} {576} (\bibinfo {year}
  {1993})}\BibitemShut {NoStop}%
\bibitem [{\citenamefont {Barhoumi}\ \emph {et~al.}(2021)\citenamefont
  {Barhoumi}, \citenamefont {Rocca}, \citenamefont {Said},\ and\ \citenamefont
  {Lebègue}}]{dimondmech}%
  \BibitemOpen
  \bibfield  {author} {\bibinfo {author} {\bibfnamefont {M.}~\bibnamefont
  {Barhoumi}}, \bibinfo {author} {\bibfnamefont {D.}~\bibnamefont {Rocca}},
  \bibinfo {author} {\bibfnamefont {M.}~\bibnamefont {Said}},\ and\ \bibinfo
  {author} {\bibfnamefont {S.}~\bibnamefont {Lebègue}},\ }\bibfield  {title}
  {\bibinfo {title} {Elastic and mechanical properties of cubic diamond and
  silicon using density functional theory and the random phase approximation},\
  }\href {https://doi.org/https://doi.org/10.1016/j.ssc.2020.114136} {\bibfield
   {journal} {\bibinfo  {journal} {Solid State Commun.}\ }\textbf {\bibinfo
  {volume} {324}},\ \bibinfo {pages} {114136} (\bibinfo {year}
  {2021})}\BibitemShut {NoStop}%
\bibitem [{\citenamefont {Popov}\ \emph {et~al.}(2013)\citenamefont {Popov},
  \citenamefont {Yang},\ and\ \citenamefont {Dunsch}}]{endo}%
  \BibitemOpen
  \bibfield  {author} {\bibinfo {author} {\bibfnamefont {A.}~\bibnamefont
  {Popov}}, \bibinfo {author} {\bibfnamefont {S.}~\bibnamefont {Yang}},\ and\
  \bibinfo {author} {\bibfnamefont {L.}~\bibnamefont {Dunsch}},\ }\bibfield
  {title} {\bibinfo {title} {Endohedral fullerenes},\ }\href
  {https://doi.org/https://doi.org/10.1021/cr300297r} {\bibfield  {journal}
  {\bibinfo  {journal} {Chem. Rev.}\ }\textbf {\bibinfo {volume} {113}},\
  \bibinfo {pages} {5989} (\bibinfo {year} {2013})}\BibitemShut {NoStop}%
\bibitem [{\citenamefont {Sato}\ \emph {et~al.}(2015)\citenamefont {Sato},
  \citenamefont {Terauchi},\ and\ \citenamefont {Yamanaka}}]{yamaka_sp3}%
  \BibitemOpen
  \bibfield  {author} {\bibinfo {author} {\bibfnamefont {Y.}~\bibnamefont
  {Sato}}, \bibinfo {author} {\bibfnamefont {M.}~\bibnamefont {Terauchi}},\
  and\ \bibinfo {author} {\bibfnamefont {S.}~\bibnamefont {Yamanaka}},\
  }\bibfield  {title} {\bibinfo {title} {Electronic structures of
  three-dimensional $\mathrm{C}_{60}$ polymers studied by
  high-energy-resolution electron energy-loss spectroscopy based on
  transmission electron microscopy},\ }\href
  {https://doi.org/https://doi.org/10.1016/j.cplett.2015.03.017} {\bibfield
  {journal} {\bibinfo  {journal} {Chem. Phys. Lett.}\ }\textbf {\bibinfo
  {volume} {626}},\ \bibinfo {pages} {90} (\bibinfo {year} {2015})}\BibitemShut
  {NoStop}%
\bibitem [{\citenamefont {Lyapin}\ \emph {et~al.}(2019)\citenamefont {Lyapin},
  \citenamefont {Katayama},\ and\ \citenamefont {Brazhkin}}]{lyapin2019}%
  \BibitemOpen
  \bibfield  {author} {\bibinfo {author} {\bibfnamefont {A.}~\bibnamefont
  {Lyapin}}, \bibinfo {author} {\bibfnamefont {Y.}~\bibnamefont {Katayama}},\
  and\ \bibinfo {author} {\bibfnamefont {V.}~\bibnamefont {Brazhkin}},\
  }\bibfield  {title} {\bibinfo {title} {Order versus disorder: In situ
  high-pressure structural study of highly polymerized three-dimensional
  $\mathrm{C}_{60}$ fullerite},\ }\href {https://doi.org/10.1063/1.5111370}
  {\bibfield  {journal} {\bibinfo  {journal} {J. Appl. Phys.}\ }\textbf
  {\bibinfo {volume} {126}},\ \bibinfo {pages} {065102} (\bibinfo {year}
  {2019})}\BibitemShut {NoStop}%
\bibitem [{\citenamefont {Zhang}\ \emph {et~al.}(2020)\citenamefont {Zhang},
  \citenamefont {Li}, \citenamefont {Luo}, \citenamefont {He}, \citenamefont
  {Gao}, \citenamefont {Soldatov}, \citenamefont {Benavides}, \citenamefont
  {Shi}, \citenamefont {Nie}, \citenamefont {Zhang}, \citenamefont {Hu},
  \citenamefont {Ma}, \citenamefont {Liu}, \citenamefont {Wen}, \citenamefont
  {Gao}, \citenamefont {Liu}, \citenamefont {Zhang}, \citenamefont {Yu},
  \citenamefont {Zhou}, \citenamefont {Zhao}, \citenamefont {Xu}, \citenamefont
  {Su}, \citenamefont {Yang}, \citenamefont {Chernogorova},\ and\ \citenamefont
  {Tian}}]{soldatov2020}%
  \BibitemOpen
  \bibfield  {author} {\bibinfo {author} {\bibfnamefont {S.}~\bibnamefont
  {Zhang}}, \bibinfo {author} {\bibfnamefont {Z.}~\bibnamefont {Li}}, \bibinfo
  {author} {\bibfnamefont {K.}~\bibnamefont {Luo}}, \bibinfo {author}
  {\bibfnamefont {J.}~\bibnamefont {He}}, \bibinfo {author} {\bibfnamefont
  {Y.}~\bibnamefont {Gao}}, \bibinfo {author} {\bibfnamefont {A.~V.}\
  \bibnamefont {Soldatov}}, \bibinfo {author} {\bibfnamefont {V.}~\bibnamefont
  {Benavides}}, \bibinfo {author} {\bibfnamefont {K.}~\bibnamefont {Shi}},
  \bibinfo {author} {\bibfnamefont {A.}~\bibnamefont {Nie}}, \bibinfo {author}
  {\bibfnamefont {B.}~\bibnamefont {Zhang}}, \bibinfo {author} {\bibfnamefont
  {W.}~\bibnamefont {Hu}}, \bibinfo {author} {\bibfnamefont {M.}~\bibnamefont
  {Ma}}, \bibinfo {author} {\bibfnamefont {Y.}~\bibnamefont {Liu}}, \bibinfo
  {author} {\bibfnamefont {B.}~\bibnamefont {Wen}}, \bibinfo {author}
  {\bibfnamefont {G.}~\bibnamefont {Gao}}, \bibinfo {author} {\bibfnamefont
  {B.}~\bibnamefont {Liu}}, \bibinfo {author} {\bibfnamefont {Y.}~\bibnamefont
  {Zhang}}, \bibinfo {author} {\bibfnamefont {D.}~\bibnamefont {Yu}}, \bibinfo
  {author} {\bibfnamefont {X.-F.}\ \bibnamefont {Zhou}}, \bibinfo {author}
  {\bibfnamefont {Z.}~\bibnamefont {Zhao}}, \bibinfo {author} {\bibfnamefont
  {B.}~\bibnamefont {Xu}}, \bibinfo {author} {\bibfnamefont {L.}~\bibnamefont
  {Su}}, \bibinfo {author} {\bibfnamefont {G.}~\bibnamefont {Yang}}, \bibinfo
  {author} {\bibfnamefont {O.}~\bibnamefont {Chernogorova}},\ and\ \bibinfo
  {author} {\bibfnamefont {Y.}~\bibnamefont {Tian}},\ }\bibfield  {title}
  {\bibinfo {title} {Discovery of carbon-based strongest and hardest amorphous
  material},\ }\href {arXiv:2011.14819} {\bibfield  {journal} {\bibinfo
  {journal} {arxiv}\ ,\ \bibinfo {pages} {40}} (\bibinfo {year}
  {2020})}\BibitemShut {NoStop}%
\bibitem [{\citenamefont {Mezouar}\ \emph {et~al.}(2003)\citenamefont
  {Mezouar}, \citenamefont {Marques}, \citenamefont {Hodeau}, \citenamefont
  {Pischedda},\ and\ \citenamefont
  {N\'u\~nez$\mathrm{-}$Regueiro}}]{mezouarprb2003}%
  \BibitemOpen
  \bibfield  {author} {\bibinfo {author} {\bibfnamefont {M.}~\bibnamefont
  {Mezouar}}, \bibinfo {author} {\bibfnamefont {L.}~\bibnamefont {Marques}},
  \bibinfo {author} {\bibfnamefont {J.-L.}\ \bibnamefont {Hodeau}}, \bibinfo
  {author} {\bibfnamefont {V.}~\bibnamefont {Pischedda}},\ and\ \bibinfo
  {author} {\bibfnamefont {M.}~\bibnamefont {N\'u\~nez$\mathrm{-}$Regueiro}},\
  }\bibfield  {title} {\bibinfo {title} {Equation of state of an anisotropic
  three-dimensional ${\mathrm{c}}_{60}$ polymer: The most stable form of
  fullerene},\ }\href {https://doi.org/10.1103/PhysRevB.68.193414} {\bibfield
  {journal} {\bibinfo  {journal} {Phys. Rev. B}\ }\textbf {\bibinfo {volume}
  {68}},\ \bibinfo {pages} {193414} (\bibinfo {year} {2003})}\BibitemShut
  {NoStop}%
\bibitem [{\citenamefont {Pei}\ and\ \citenamefont {Wang}(2019)}]{reviewpei}%
  \BibitemOpen
  \bibfield  {author} {\bibinfo {author} {\bibfnamefont {C.}~\bibnamefont
  {Pei}}\ and\ \bibinfo {author} {\bibfnamefont {L.}~\bibnamefont {Wang}},\
  }\bibfield  {title} {\bibinfo {title} {Recent progress on high-pressure and
  high-temperature studies of fullerenes and related materials},\ }\href
  {https://doi.org/10.1063/1.5086310} {\bibfield  {journal} {\bibinfo
  {journal} {Matter Radiat. at Extremes}\ }\textbf {\bibinfo {volume} {4}},\
  \bibinfo {pages} {028201} (\bibinfo {year} {2019})}\BibitemShut {NoStop}%
\bibitem [{\citenamefont {Kresse}\ and\ \citenamefont
  {Furthm\"uller}(1996)}]{i32}%
  \BibitemOpen
  \bibfield  {author} {\bibinfo {author} {\bibfnamefont {G.}~\bibnamefont
  {Kresse}}\ and\ \bibinfo {author} {\bibfnamefont {J.}~\bibnamefont
  {Furthm\"uller}},\ }\bibfield  {title} {\bibinfo {title} {Efficient iterative
  schemes for ab initio total-energy calculations using a plane-wave basis
  set},\ }\href {https://doi.org/10.1103/PhysRevB.54.11169} {\bibfield
  {journal} {\bibinfo  {journal} {Phys. Rev. B}\ }\textbf {\bibinfo {volume}
  {54}},\ \bibinfo {pages} {11169} (\bibinfo {year} {1996})}\BibitemShut
  {NoStop}%
\bibitem [{\citenamefont {Perdew}\ \emph {et~al.}(1996)\citenamefont {Perdew},
  \citenamefont {Burke},\ and\ \citenamefont {Ernzerhof}}]{i2}%
  \BibitemOpen
  \bibfield  {author} {\bibinfo {author} {\bibfnamefont {J.}~\bibnamefont
  {Perdew}}, \bibinfo {author} {\bibfnamefont {K.}~\bibnamefont {Burke}},\ and\
  \bibinfo {author} {\bibfnamefont {M.}~\bibnamefont {Ernzerhof}},\ }\bibfield
  {title} {\bibinfo {title} {Generalized gradient approximation made simple},\
  }\href {https://doi.org/10.1103/PhysRevLett.77.3865} {\bibfield  {journal}
  {\bibinfo  {journal} {Phys. Rev. Lett.}\ }\textbf {\bibinfo {volume} {77}},\
  \bibinfo {pages} {3865} (\bibinfo {year} {1996})}\BibitemShut {NoStop}%
\bibitem [{\citenamefont {Wu}\ \emph {et~al.}(2005)\citenamefont {Wu},
  \citenamefont {Vanderbilt},\ and\ \citenamefont {Hamann}}]{vasp_stresstrain}%
  \BibitemOpen
  \bibfield  {author} {\bibinfo {author} {\bibfnamefont {X.}~\bibnamefont
  {Wu}}, \bibinfo {author} {\bibfnamefont {D.}~\bibnamefont {Vanderbilt}},\
  and\ \bibinfo {author} {\bibfnamefont {D.}~\bibnamefont {Hamann}},\
  }\bibfield  {title} {\bibinfo {title} {Systematic treatment of displacements,
  strains, and electric fields in density-functional perturbation theory},\
  }\href {https://doi.org/10.1103/PhysRevB.72.035105} {\bibfield  {journal}
  {\bibinfo  {journal} {Phys. Rev. B}\ }\textbf {\bibinfo {volume} {72}},\
  \bibinfo {pages} {035105} (\bibinfo {year} {2005})}\BibitemShut {NoStop}%
\bibitem [{\citenamefont {Singh}\ \emph {et~al.}(2018)\citenamefont {Singh},
  \citenamefont {Valencia-Jaime}, \citenamefont {Pavlic},\ and\ \citenamefont
  {Romero}}]{mechelastic}%
  \BibitemOpen
  \bibfield  {author} {\bibinfo {author} {\bibfnamefont {S.}~\bibnamefont
  {Singh}}, \bibinfo {author} {\bibfnamefont {I.}~\bibnamefont
  {Valencia-Jaime}}, \bibinfo {author} {\bibfnamefont {O.}~\bibnamefont
  {Pavlic}},\ and\ \bibinfo {author} {\bibfnamefont {A.}~\bibnamefont
  {Romero}},\ }\bibfield  {title} {\bibinfo {title} {Elastic, mechanical, and
  thermodynamic properties of $\mathrm{Bi-Sb}$ binaries: Effect of spin-orbit
  coupling},\ }\href {https://doi.org/10.1103/PhysRevB.97.054108} {\bibfield
  {journal} {\bibinfo  {journal} {Phys. Rev. B}\ }\textbf {\bibinfo {volume}
  {97}},\ \bibinfo {pages} {054108} (\bibinfo {year} {2018})}\BibitemShut
  {NoStop}%
\bibitem [{\citenamefont {Togo}\ and\ \citenamefont {Tanaka}(2015)}]{phonopy}%
  \BibitemOpen
  \bibfield  {author} {\bibinfo {author} {\bibfnamefont {A.}~\bibnamefont
  {Togo}}\ and\ \bibinfo {author} {\bibfnamefont {I.}~\bibnamefont {Tanaka}},\
  }\bibfield  {title} {\bibinfo {title} {First principles phonon calculations
  in materials science},\ }\href
  {https://doi.org/https://doi.org/10.1016/j.scriptamat.2015.07.021} {\bibfield
   {journal} {\bibinfo  {journal} {Scr. Mater.}\ }\textbf {\bibinfo {volume}
  {108}},\ \bibinfo {pages} {1} (\bibinfo {year} {2015})}\BibitemShut {NoStop}%
\bibitem [{\citenamefont {Barone}\ \emph {et~al.}(2011)\citenamefont {Barone},
  \citenamefont {Hod}, \citenamefont {Peralta},\ and\ \citenamefont
  {Scuseria}}]{hse}%
  \BibitemOpen
  \bibfield  {author} {\bibinfo {author} {\bibfnamefont {V.}~\bibnamefont
  {Barone}}, \bibinfo {author} {\bibfnamefont {O.}~\bibnamefont {Hod}},
  \bibinfo {author} {\bibfnamefont {J.}~\bibnamefont {Peralta}},\ and\ \bibinfo
  {author} {\bibfnamefont {G.}~\bibnamefont {Scuseria}},\ }\bibfield  {title}
  {\bibinfo {title} {Accurate prediction of the electronic properties of
  low-dimensional graphene derivatives using a screened hybrid density
  functional},\ }\href {https://doi.org/10.1021/ar100137c} {\bibfield
  {journal} {\bibinfo  {journal} {Acc. Chem. Res.}\ }\textbf {\bibinfo {volume}
  {44}},\ \bibinfo {pages} {269} (\bibinfo {year} {2011})}\BibitemShut
  {NoStop}%
\bibitem [{\citenamefont {Frisch}\ and\ \citenamefont {et~al}(2016)}]{g09}%
  \BibitemOpen
  \bibfield  {author} {\bibinfo {author} {\bibfnamefont {M.}~\bibnamefont
  {Frisch}}\ and\ \bibinfo {author} {\bibnamefont {et~al}},\ }\href@noop {}
  {\bibinfo {title} {Gaussian 09 revision {A}.02}} (\bibinfo {year} {2016}),\
  \bibinfo {note} {gaussian Inc. Wallingford CT, 2016}\BibitemShut {NoStop}%
\bibitem [{\citenamefont {Mouhat}\ and\ \citenamefont
  {Coudert}(2014)}]{elasticcretiria}%
  \BibitemOpen
  \bibfield  {author} {\bibinfo {author} {\bibfnamefont {F.}~\bibnamefont
  {Mouhat}}\ and\ \bibinfo {author} {\bibfnamefont {F.-X.}\ \bibnamefont
  {Coudert}},\ }\bibfield  {title} {\bibinfo {title} {Necessary and sufficient
  elastic stability conditions in various crystal systems},\ }\href
  {https://doi.org/10.1103/PhysRevB.90.224104} {\bibfield  {journal} {\bibinfo
  {journal} {Phys. Rev. B}\ }\textbf {\bibinfo {volume} {90}},\ \bibinfo
  {pages} {224104} (\bibinfo {year} {2014})}\BibitemShut {NoStop}%
\bibitem [{\citenamefont {Hill}(1952)}]{Hill_1952}%
  \BibitemOpen
  \bibfield  {author} {\bibinfo {author} {\bibfnamefont {R.}~\bibnamefont
  {Hill}},\ }\bibfield  {title} {\bibinfo {title} {The elastic behaviour of a
  crystalline aggregate},\ }\href {https://doi.org/10.1088/0370-1298/65/5/307}
  {\bibfield  {journal} {\bibinfo  {journal} {Proc. Phys. Soc. Section A}\
  }\textbf {\bibinfo {volume} {65}},\ \bibinfo {pages} {349} (\bibinfo {year}
  {1952})}\BibitemShut {NoStop}%
\bibitem [{\citenamefont {Chen}\ \emph {et~al.}(2011)\citenamefont {Chen},
  \citenamefont {Niu}, \citenamefont {Li},\ and\ \citenamefont
  {Li}}]{chenhardness}%
  \BibitemOpen
  \bibfield  {author} {\bibinfo {author} {\bibfnamefont {X.-Q.}\ \bibnamefont
  {Chen}}, \bibinfo {author} {\bibfnamefont {H.}~\bibnamefont {Niu}}, \bibinfo
  {author} {\bibfnamefont {D.}~\bibnamefont {Li}},\ and\ \bibinfo {author}
  {\bibfnamefont {Y.}~\bibnamefont {Li}},\ }\bibfield  {title} {\bibinfo
  {title} {Modeling hardness of polycrystalline materials and bulk metallic
  glasses},\ }\href
  {https://doi.org/https://doi.org/10.1016/j.intermet.2011.03.026} {\bibfield
  {journal} {\bibinfo  {journal} {Intermetallics}\ }\textbf {\bibinfo {volume}
  {19}},\ \bibinfo {pages} {1275} (\bibinfo {year} {2011})}\BibitemShut
  {NoStop}%
\end{thebibliography}%

\cleardoublepage

\newpage

\begin{figure*}[t]
	\centering
\includegraphics[width=\textwidth,page=1]{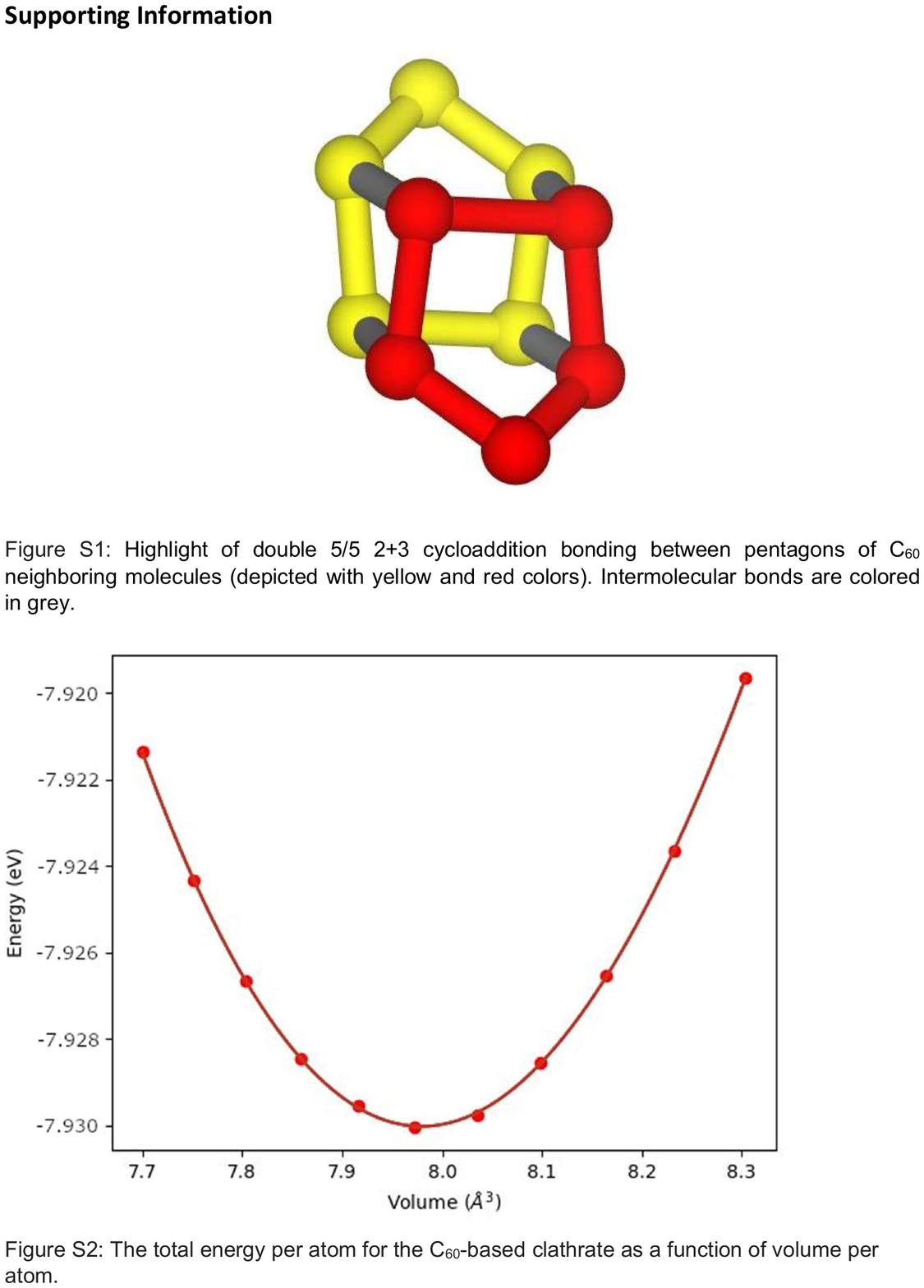}
\end{figure*}

\begin{figure*}[t]
	\centering
\includegraphics[width=\textwidth,page=2]{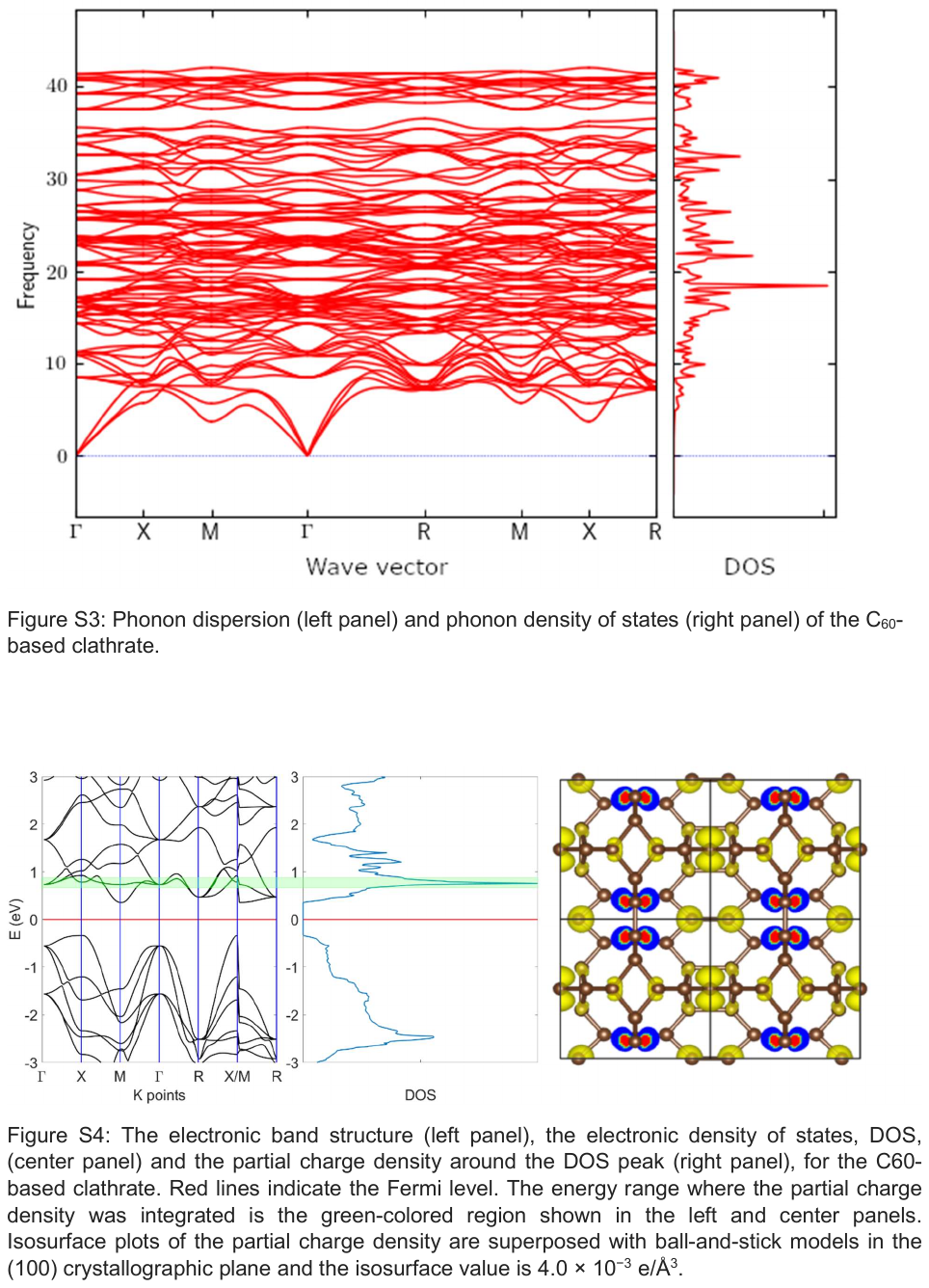}
\end{figure*}

\begin{figure*}[t]
	\centering
\includegraphics[width=\textwidth,page=3]{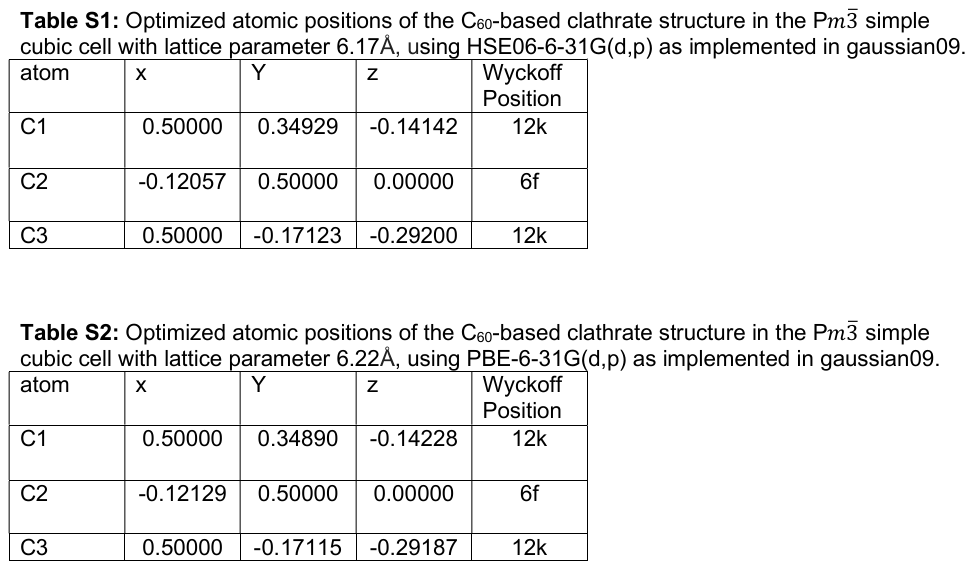}
\end{figure*}

\end{document}